# A sol-gel method for growing superconducting MgB$_2$ films

Liping Chen[1], Chen Zhang[1], Yinbo Wang[1], Yue Wang[1,*], Qingrong Feng[1,†], Zizhao Gan[1],

Junzhi Yang[2] and Xingguo Li[2]

[1] *State Key Laboratory for Mesoscopic Physics, Applied Superconductivity Center, and School of Physics, Peking University, Beijing 100871, People's Republic of China*

[2] *College of Chemistry and Molecular Engineering, Peking University, Beijing 100871, People's Republic of China*

[*] yue.wang@pku.edu.cn

[†] qrfeng@pku.edu.cn

## Abstract

In this paper we report a new sol-gel method for the fabrication of MgB$_2$ films. Polycrystalline MgB$_2$ films were prepared by spin-coating a precursor solution of Mg(BH$_4$)$_2$ diethyl ether on (001)Al$_2$O$_3$ substrates followed with annealing in Mg vapor. In comparison with the MgB$_2$ films grown by other techniques, our films show medium qualities including a superconducting transition temperature of $T_c \sim 37$ K, a critical current density of $J_c$(5 K, 0 T) $\sim 5 \times 10^6$ A·cm$^{-2}$, and a critical field of $H_{c2}(0) \sim 19$ T. Such a sol-gel technique shows potential in the commercial fabrication of practically used MgB$_2$ films as well as MgB$_2$ wires and tapes.

The discovery of superconductivity in binary metallic $MgB_2$ has stimulated great interest in the studies of its superconducting mechanism, physical properties and possible applications [1, 2]. $MgB_2$ presents a relatively high superconducting transition temperature of 39 K, metallic charge carrier density and intergrains of a strongly linked nature. These properties make it a promising replacement for the existing 'low $T_c$' materials as well as the 'high $T_c$' superconductors in electronics and power applications. $MgB_2$-coated conductors are essential to some devices such as the superconducting particle accelerator. Consequently, several techniques to fabricate the $MgB_2$-coated conductors have been developed. For instance, hybrid physical-chemical vapor deposition (HPCVD) [3-7] and *ex situ* high temperature annealing in Mg vapor [8-15] have been reported. The cost consideration in commercial applications highlights a cheap film making method. The electrophoresis method was ever used to deposit precursor B films in *ex situ* fabrication of $MgB_2$ thick films [15]. Recently, the co-deposition method has also been reported by Moeckly and Ruby [16]. This method allows large-area $MgB_2$ film with high qualities, and is prospective to be a low cost method of making films. In the synthesis of 'high $T_c$' superconducting thick films, various methods were used, such as the sol-gel [17-19], the partial melting technique [20] and so on. The sol-gel technique has the advantages of low cost and ease of obtaining homogeneous coatings on any curved surface with large areas. The sol-gel synthesis of superconducting $MgB_2$ nanowire has been studied [21]. However, a sol-gel technique for the fabrication of $MgB_2$ films has never been reported. As a material for reserving hydrogen, $Mg(BH_4)_2$ has been previously investigated, and presents a chemical reaction of $Mg(BH_4)_2 \rightarrow MgB_2 + 4H_2$ at temperatures near 580 °C [22]. Such behavior implied $Mg(BH_4)_2$ could be utilized as a potential material for the fabrication of $MgB_2$. If a proper solution of $Mg(BH_4)_2$ can be obtained, a sol-gel method for the deposition of $MgB_2$-coated conductors is possible. In this paper, we fabricated $MgB_2$ films in a sol-gel method by using $Mg(BH_4)_2$ diethyl ether solution ($Mg(BH_4)_2 \cdot Et_2O$).

The $Mg(BH_4)_2 \cdot Et_2O$ precursor solution was prepared by milling $NaBH_4$ and $MgCl_2$ followed by refluxing in diethyl ether with the reaction of $2NaBH_4 + MgCl_2 \rightarrow Mg(BH_4)_2 + 2NaCl$ [23]. The NaCl was removed from the solution due to its indiscerptibility in diethyl ether. In order to prevent possible oxidation, both the processes of milling and refluxing were carried out under an argon or hydrogen atmosphere. We fabricated $MgB_2$ films by spin-coating a $Mg(BH_4)_2 \cdot Et_2O$ solution on $Al_2O_3$ substrates and post-annealing in Mg vapor at 700 °C for 15 min. The post-annealing was performed in a quartz tube with a $H_2$ atmosphere. Before annealing in Mg Vapor, the samples were heated up to 140 °C and held at this temperature for 10 min to evaporate the diethyl ether sufficiently. In order to prevent cracking, the temperature was controlled to be changed slowly. Previous studies implied that $Mg(BH_4)_2$ was decomposed to $MgB_2$ at the temperature of 580 °C; when the temperature was lower than 410 °C, metal Mg would be decomposed from $Mg(BH_4)_2$ [22]. Considering the pyrolysis of $Mg(BH_4)_2$ into $MgB_2$ and evaporation of Mg ingots, we used the anneal temperature at 700 °C. The Mg vapor was used to prevent the deficiency of Mg in the reaction. The films' thicknesses were controlled by the time of spin-coating, and in present study they ranged between

200 and 1000 nm.

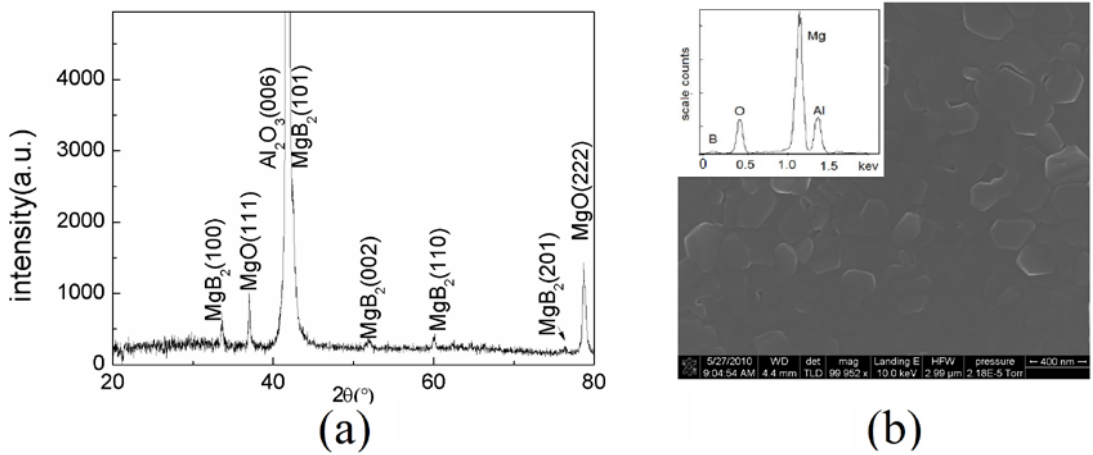

**Figure 1.** (a) The XRD pattern of the $MgB_2$ films on sapphire substrates deposited by the sol-gel method. (b) SEM image of the $MgB_2$ film magnified by 100 thousand. The inset shows the qualitative EDS of the film.

The growth of the films was explored by x-ray diffraction (XRD) and scanning electric micrograph-energy dispersive x-ray spectroscopy (SEM-EDS). The XRD pattern of one typical film is shown in figure 1(a). Besides the reflection of the (001)-oriented $Al_2O_3$ substrate, polycrystalline $MgB_2$ phase with impurities of MgO can be indexed. For MgO, it is interesting to note that only (111) and (222) peaks are observed, which seems to suggest a [111]-oriented growth manner. It is known that MgO has a cubic lattice structure and along its [111] direction there is a hexagonal lattice symmetry. Hence, with the $Al_2O_3$ substrate having a similar hexagonal lattice structure, a stacking of MgO along its [111] direction may be preferred because of the better lattice match. The exposure to air during the spin coating, the low vacuum of quartz tube and the impurity in Mg ingots may be responsible for the appearance of the MgO impurity in our films. Solution coating in an Ar atmosphere and *ex situ* annealing in high purity Mg vapor are expected to improve the purity of the film. The absence of $Mg(BH_4)_2$ in XRD shows that its pyrolysis is complete. No obvious NaCl or $MgCl_2$ phase is observed in the XRD pattern. This result indicates that the $NaBH_4$ reacted with the $MgCl_2$ completely, and the NaCl was fully removed from the precursor solution of $Mg(BH_4)_2 \cdot Et_2O$. The absence of NaCl or $MgCl_2$ was also confirmed by the energy dispersive x-ray spectroscopy (EDS) analysis. The qualitative EDS data is present in the inset of figure 1(b). Apart from B, O, Mg and Al, no other element such as Na, Cl is observed, indicating the purity of the $Mg(BH_4)_2 \cdot Et_2O$ precursor. Figure 1(b) shows the SEM image of the film magnified by 100 thousand. Hexagonal $MgB_2$ crystal grains with a size of 200 nm are aligned densely, resulting in a relatively smooth surface without obvious cracks. The films are difficult to remove from the substrate by mechanical force, showing strong adhesion between the films and the substrates.

We measured the resistivity dependence on temperature by using a standard four-probe method with the Quantum Design PPMS System. Figure 2 shows the temperature dependence of resistivity of the film. The inset is a magnified view near the transition region. The film begins superconducting transition at a temperature of 37.4 K, and completely gets into a zero resistive state at 35.1 K, showing a transition width of $\Delta T \sim 2$ K. The resistivity at 300 K and 40 K are 28 $\mu\Omega$·cm and 15.8 $\mu\Omega$·cm, respectively, which gives a residual resistivity ratio RRR($\rho_{300 K}/\rho_{40 K}$) of about 1.8. Although bettered by those of films fabricated by HPCVD [3-7] and co-deposition methods [16], these values of our samples are comparable to most films prepared by post-annealing in Mg vapor [8-15]. Using the formalism proposed by Rowell [24], we can tentatively estimate the fractional area $A_F$ of the sample that carries current from the equation $A_F = \Delta\rho_{ideal}/(\rho_{300 K} - \rho_{40 K})$ where $\rho_T$ is our experimentally measured resistivity at temperature $T$ and $\Delta\rho_{ideal}$ is the corresponding resistivity change in the single crystal. With $\Delta\rho_{ideal} = 4.3$ $\mu\Omega$·cm [25], we find the $A_F$ of the film is about 35%, considerably lower than the unity. This somehow lower connectivity may be due to the presence of MgO impurities in the film. As indicated above by the XRD measurement, the MgO within the MgB$_2$ film matrix may reduce the active cross-sectional area of the sample.

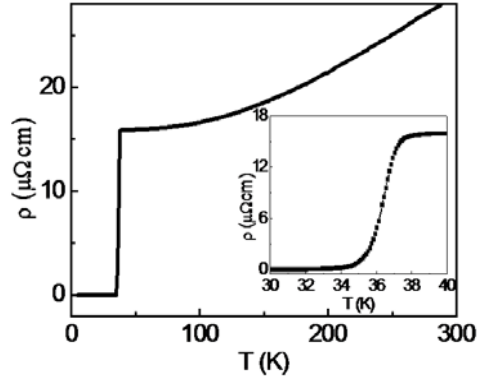

**Figure 2.** The typical resistivity versus temperature of a film; the inset is the magnified view near the $T_c$ region.

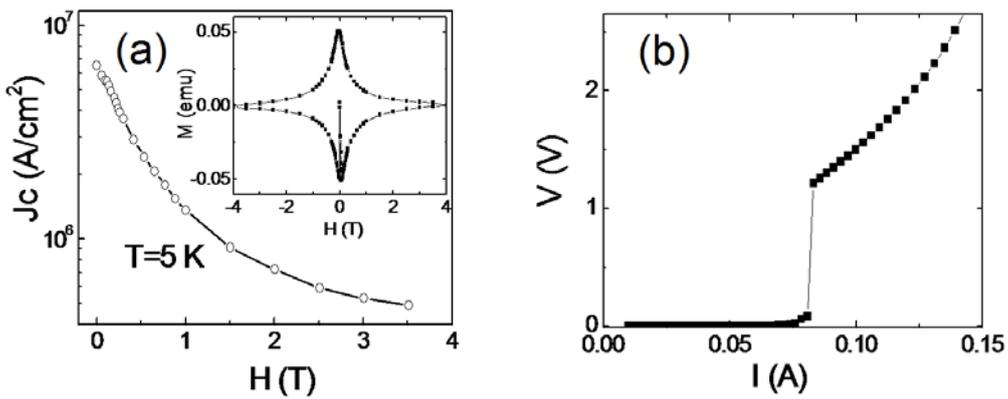

**Figure 3.** (a) The field dependence of the critical current density $J_c$ estimated using the Bean model. The inset is the *M-H* loop of the film measured at a temperature of 5 K. (b) The voltage-current (*V-I*) curve of a 10 $\mu$m wide line patterned in a film measured at 5 K.

The *M-H* measurements were performed at 5 K on the Quantum Design MPMS-7. The typical *M-H* loop of a film with 200 nm thickness is present in the inset of figure 3(a). The critical current density is obtained from the *M-H* loops using the Bean critical model $J_c = 20\Delta M/[Va(1-a/3b)]$ ($a < b$), where $\Delta M$ is the height of the *M-H* loops, $a$ and $b$ are the lateral dimensions of the films, and $V$ is the volume of the films. In our experiment, the film is patterned by 1 mm × 1.8 mm. Figure 3(a) exhibits the field dependence of the critical current density $J_c$. $J_c$(5 K, 0 T) is $6.5 \times 10^6$ A·cm$^{-2}$ while $J_c$(5 K, 3 T) is lowered to $5 \times 10^5$ A·cm$^{-2}$. We can see that $J_c$ drops significantly with increasing magnetic field. One possible explanation for such behavior of $J_c$ may relate to the presence of MgO impurity in the film. As discussed above, MgO impurity appearing in the film may be a major detrimental factor to the connectivity of the sample. It was also shown by others that the intergranular MgO was an important current barrier in MgB$_2$ and could cause a strong depression of $J_c$ [26]. As the magnetic field increases, the deleterious effect of MgO on the $J_c$ may become larger and hence lead to the rapid reduction of the *M-H* loop. In addition to the *M-H* measurement, we also determined the $J_c$ of the film by transport measurement at a temperature of 5 K. Figure 3(b) shows the voltage-current (*V-I*) curve obtained on a 100 μm long, 10 μm wide line patterned in a 500 nm thick film. We can see the voltage shows an apparent jump at $I \sim 80$ mA and roughly linearly increases with further increasing the current. This suggests the critical current of the sample to be about 80 mA. As a result, a critical current density $J_c$ of about $1.6 \times 10^6$ A·cm$^{-2}$ was evaluated. This transport $J_c$ is somehow lower than the above estimated from the magnetization measurement but both have the same magnitude.

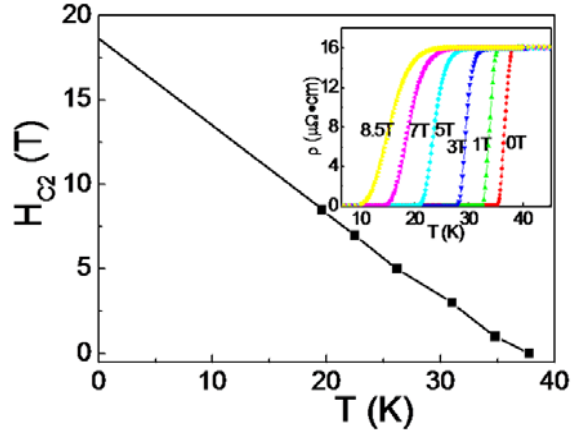

**Figure 4.** The critical field $H_{c2}$ against the temperature corresponding to the 90% points on the resistive transition. The inset shows the resistivity as a function of temperature in various fields from 0 to 8.5 T.

The upper critical field $H_{c2}$ is another important factor in the potential applications of MgB$_2$. To determine $H_{c2}$, we measured the resistivity as a function of the temperature in various fields from 0 to 8.5 T, as shown in the inset of figure 4. With increasing filed, it is seen that the transition moves to a lower temperature steadily. By choosing the point at which the resistivity becomes 90% of the value just before the transition, we obtained $H_{c2}(T)$ which we plotted in the main panel of figure 4. A roughly linear behavior of $H_{c2}(T)$ in $T$ with a slope of -0.45 is observed. The estimated $H_{c2}$ at 0 K is about 19 T which is higher than that reported for the pure $MgB_2$ [3]. Some reports suggested that for $MgB_2$ the $H_{c2}$ may depend on the resistivity of the sample and be enhanced by impurity scattering [27, 28]. According to this proposal, the relatively high $H_{c2}$ of our film could be linked with the relatively large residual resistivity of the sample, which in turn may arise from the scattering within the $MgB_2$ grains because of the MgO impurity or disorder effect. For the $MgB_2$ films fabricated by post-annealing in Mg vapor, it was reported that $H_{c2}$ depended on details of the annealing such as the annealing temperature [29]. With this we speculate that an even higher $H_{c2}$ of $MgB_2$ films fabricated by the sol-gel method might be obtained by adjusting the annealing conditions.

Collectively, we have shown that the films fabricated by the sol-gel method have a $T_c$ of 35-38 K, a $J_c$(5 K, 0 T) of about $5 \times 10^6$ A·cm$^{-2}$, and a $H_{c2}(0)$ of around 19 T. These values of the $T_c$, $J_c$ and $H_{c2}$ of the films are comparable to most reported values for bulk samples, and could satisfy the needs of practical applications. In addition, the sol-gel fabrication of $MgB_2$ films has some advantages. It may be the trend to overcome cost and speed considerations. In spite of outstanding superconducting properties of the *in-situ* grown thick films, poisonous and expensive $B_2H_6$ is concerned in the HPCVD method. The sources of $NaBH_4$ and $MgCl_2$ used in the sol-gel method are abundant and low cost. The sol-gel method is not only commercial but also has a high speed in $MgB_2$ film fabrication due to its simple technique. This method is also suitable for deposition of large-scale films with high reliability and reproducibility. Thus, the sol-gel deposition of $MgB_2$ films holds great potential as a fast and efficient method of producing large-scale homogeneous superconducting films at a lower cost.

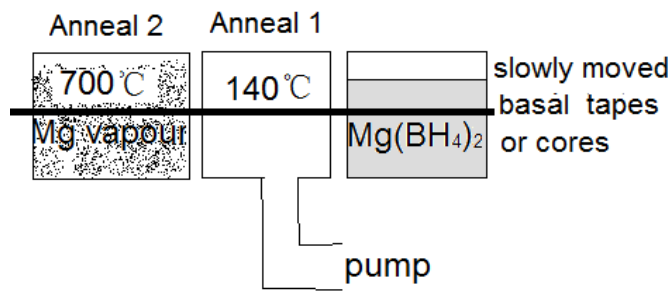

**Figure 5.** The schematic system to fabricate long $MgB_2$ tapes or wires in the sol-gel method.

The technology of sol-gel fabrication of $MgB_2$ thick films has potential for making $MgB_2$ long wires and tapes. As we know, the sol-gel technique has been widely used for the fabrication of 'high-$T_c$' superconductor films [19] as well as insulating coating in superconducting tapes or wires such as Bi-2212 tape [30], $MgB_2$ wires [31] and so on. At present, powder-in-tube (PIT) technique is widely used in fabricating practically used $MgB_2$ tapes or wires. Ag [32], Cu [33, 34] or Fe [35, 36] sheathed $MgB_2$ wires have been fabricated by PIT, though sometimes the $MgB_2$ wires are porous. An efficient technique to fabricate decent-quality $MgB_2$ wires or tapes is still needed. The sol-gel fabrication of $MgB_2$ thick films can be spread to make $MgB_2$ long wires and tapes. By coating with a precursor solution of $Mg(BH_4)_2$ followed with annealing in Mg vapor with slowly moved basal tapes or cores, $MgB_2$ long wires and tapes on metal substrates can be deposited. The schematic design is shown in figure 5. This work is now in progress. Considering $J_c$(5 K, 0 T) ~ $5 \times 10^6$ A·cm$^{-2}$ of the sol-gel deposited thick films, the tapes made by the sol-gel method are expected to have a critical current comparable with those fabricated by the reported PIT technology. Moreover, the sol-gel technique provides us a method to fabricated nanometer powder of $MgB_2$ which can be used in PIT. In this way, the oxidation of Mg powder during the process of the production and storage in the solid phase reaction is avoided, and thus superior qualities of wire are expected.

In conclusion, we have fabricated polycrystalline $MgB_2$ thick films on substrates of sapphire by using a sol-gel method with a precursor solution of $Mg(BH_4)_2$ diethyl ether. The obtained films with strong adhesion to substrates are comparable to most samples deposited by post-annealing in Mg vapor in terms of their superconducting properties such as transition temperature, critical current density and so on. This advance improves the commercial fabrication of $MgB_2$ thick films and even $MgB_2$ wires and tapes for industrial applications.

## Acknowledgments

This work is supported by the Ministry of Science and Technology of China (973 Program, No. 2006CD601004), and the Research Foundation for Excellent Undergraduates of the Ministry of Education of China (Grant No. J0630311).